\documentclass[aps,prd,onecolumn,groupedaddress,amssymb,showpacs,nofootinbib,epsfig,pdffig]{revtex4}
\usepackage{graphicx}
\usepackage{bm}
\usepackage{dcolumn}
\usepackage{amsmath}
\usepackage{amssymb}
\usepackage{epsfig}
\usepackage[colorlinks,linkcolor=blue,citecolor=blue,urlcolor=blue]{hyperref}
\newcommand{\be}{\begin{equation}}
\newcommand{\ee}{\end{equation}}

\begin{document}

\title{Dynamics of nonminimally coupled scalar field models with generic potentials in FLRW background }

\author{M. Shahalam$^{1,2}$ \footnote{E-mail address: shahalam@zjut.edu.cn}}
\author{Shynaray  Myrzakul$^{2,3,4}$ \footnote{E-mail address: shynaray1981@gmail.com}}
\affiliation{$^{1}$Institute for Theoretical Physics $\&$ Cosmology, Zhejiang University of Technology, Hangzhou- 310023, China}
\affiliation{$^{2}$Ratbay Myrzakulov Eurasian International Centre for Theoretical Physics, Nur-Sultan-010009, Kazakhstan}
\affiliation{$^{3}$Eurasian  National University, Nur-Sultan- 010008, Kazakhstan}
\affiliation{$^{4}$Nazarbayev University,  Nur-Sultan-010000, Kazakhstan}

\begin{abstract}
We study the phase space analysis of a nonminimally coupled scalar field model with different potentials such as KKLT, Higgs, inverse and inverse square. Our investigation brings new asymptotic regimes, and obtains stable de-Sitter solution. In case of KKLT, we do not find stable de-Sitter solution whereas Higgs model satisfies the de-Sitter condition but does not provide a stable de-Sitter solution in usual sense as one of the eigenvalue is zero. We obtain time derivative of Hubble constant $\dot{H}=0$, equation of state $w_{\phi}\simeq -1$, scalar field $\phi=$constant and the positive effective gravitational constant ($G_{eff}>0$), which are missed in our earlier work. Therefore, in case of $F(\phi)R$ coupling with $F(\phi)= 1-\xi\phi^2 $ and the models of inverse and inverse square potentials$-$ a true stable de-Sitter solution is trivially satisfied.
\end{abstract}
\maketitle

\section{Introduction}
\label{sec:intro}
In the current scenerio, the scalar fields play a vital role in cosmological investigations. Scalar fields are used in quintessence, phantom, inflation and the dynamical behavior of loop quantum cosmology (LQC) etc. \cite{review2,alamLQC}. In the context of quintessence, the energy density of the minimally coupled scalar field to gravity mimics the effective cosmological behavior. Naturally, the detailed dynamics would be dependent on the specific form of the potential function. We extend the quintessence framework by including the nonminimally coupling constant to gravity, and it is well known scalar-tensor theory that has been investigated for many years, and emerged in Brans-Dicke theory to match the Mach's principle with general theory of relativity \cite{BD}. In this theory, the Newtonian gravitational constant is a time varying function that appears into the action with the curvature term. Due to novel charectristics, the nonminimally coupled (NMC) scalar field models are of great interest to dark energy \cite{PR1,PR2,review1,vpaddy,review3,review3C,review3d,review4}, and have been extensively discussed in \cite{a1,a2,a3,a4,a5,a6,a7,a8,a9,a10,a11,a12,a13,a14,a15,a16,sunny,g1}. A well known model of NMC scalar field system is supported by $F(\phi) R$ coupling with $F(\phi)=1- \xi \phi^2$. In the literature, the NMC scalar field models have been widely used in the framework of late time cosmological behavior. For instance, nonminimal coupling avoids the coincidence problem, may allow phantom crossing and cosmological scaling solutions. The generic features of a NMC scalar field model is phantom crossing having  $F(\phi)=1- \xi \phi^2$ \cite{Polarski}.

A dynamical system theory plays a key role to understand the asymptotic behavior of various cosmological models. These models may give asymptotic solutions, and their stability can be confirmed by a simple programmed algorithm. As a result, stability and phase space trajectories provide the viable cosmological behaviors. In this paper, we consider four models with different potentials, namely,  KKLT (Kachru, Kallosh, Linde and Trivedi), Higgs, inverse and inverse square potentials. We investigate the dynamics of a NMC scalar field model having a suitable form of $F(\phi)$ with said potentials, and discuss the stationary points and their stability. We choose the functional form of  $F(\phi)$ as $F(\phi)=1- \xi B(\phi)$ with $B(\phi) \propto \phi^2$. We shall use exactly same equations of  autonomous system that has been given in Ref. \cite{alam2012}. However, in our previous work, we missed the generic features of underlying dynamics as no stable de-Sitter solution was found because the effective gravitational constant $G_{eff}$ is negative for $B(\phi) \propto \phi^N$ $(N \geq 2)$ and $V(\phi) \propto \phi^n$. Though, in the present work, the stable de-Sitter solution is trivially satisfied in case of NMC scalar field model with inverse and inverse square potentials.

Rest of the paper is organized as follows. In Section \ref{sec:EOM}, we study the evolution equations for a spatially flat Friedmann-Leimetre-Robertson-Walker (FLRW) universe containing a NMC scalar field model, and obtain the autonomous system which is useful to draw the phase portraits. In Section \ref{sec:phase}, we present phase space analysis for KKLT, Higgs, inverse and inverse square potentials. The behavior of $G_{eff}$ is discussed in Section \ref{sec:Geff}, and the results are summarized in Section \ref{sec:conc}.

\section{Equations of motion}
\label{sec:EOM}
Let us consider the following action with a NMC scalar field model \cite{alam2012,alam2020}
 \be
 \label{eq:Lagrangian}
S=\frac{1}{2}\int{\sqrt{-g}d^4x\Big{[} m_{Pl}^2
R-(g^{\mu\nu}\phi_{\mu}\phi_{\nu}+ \xi R
B(\phi)+2V(\phi))\Big{]}}+S_M, \ee
where $m_{Pl}^2=({8\pi
G})^{-1}=({\kappa})^{-1}$, $\xi$ denotes a dimensionless coupling constant and $S_M$ is the matter action.

The evolution equations in a spatially flat FLRW background are obtained by varying the action (\ref{eq:Lagrangian}), and given by
\be \label{eq:Friedphi} H^2=\frac{\kappa}{3}\left(\frac{1}{2}{\dot
{\phi}}^{2}+V(\phi)+3\xi(H \dot {\phi} B'(\phi)+H^{2}B(\phi))+\rho
\right), \ee
\begin{eqnarray}
\label{eq:Friedphi2} R=\kappa \left(-{\dot {\phi}}^{2} +4
V(\phi)+3\xi(3 H \dot{\phi}  B'(\phi)+\frac{R}{3}  B(\phi)  + {\dot
{\phi}}^{2}B''(\phi)+\ddot {\phi} B'(\phi))+\rho(1-3\omega) \right),
\end{eqnarray}
\begin{eqnarray}
\label{eq:KGphi}
&&\ddot {\phi}+3 H \dot {\phi}+\frac{1}{2}\xi R B'(\phi)+V'(\phi)=0.
\end{eqnarray}
Where $p$ and $\rho$  designate the pressure and energy density of the matter, respectively.

From the standard form of equations
\begin{eqnarray*}
R_{ij}-\frac{1}{2}R g_{ij}&=&8\pi G_{eff}(T_{ij,\phi}+T_{ij,m})=\kappa
T^{eff}_{ij}.
\end{eqnarray*}
The effective Newtonian gravitational constant can be written as \cite{alam2012}
\begin{equation}
G_{eff}=\frac{\kappa}{8\pi(1-\kappa \xi B(\phi))}.
\end{equation}
We also define Ricci Scalar as $R=6(2H^2+ \dot {H})$. We choose $\kappa=6$ for simplicity \cite{alam2012}.

Dividing equations (\ref{eq:Friedphi}), (\ref{eq:Friedphi2}), (\ref{eq:KGphi})
by $H^2(1-6\xi B(\phi))$ and  multiplying equation (\ref{eq:KGphi}) by $\xi
B'(\phi)$, we have
\begin{eqnarray}
\label{eq:newFried1} 1=\frac{{\dot {\phi}}^{2} }{H^2(1-6\xi
B(\phi))}+\frac{2 V(\phi)}{H^2(1-6\xi B(\phi))} +\frac{6\xi \dot
{\phi} B'(\phi)}{H(1-6\xi B(\phi))}+\frac{2\rho}{H^2(1-6\xi
B(\phi))},
\end{eqnarray}
\begin{eqnarray}
\label{eq:newFried2} \frac{R}{H^2}&=&-\frac{6 {\dot {\phi}}
^{2}}{H^2(1-6 \xi B(\phi))}+\frac{24 V(\phi)}{H^2(1-6 \xi B(\phi))}
+\frac{54 \xi \dot {\phi} B'(\phi)}{H(1-6\xi B(\phi))}+\frac{18 \xi
{\dot{\phi}} ^2 B''(\phi)}{H^2(1-6 \xi B(\phi))}\nonumber\\
&+&\frac{18 \xi \ddot {\phi}  B'(\phi)}{H^2(1-6 \xi B(\phi))}
 +\frac{6\rho(1-3\omega)}{H^2(1-6 \xi B(\phi))},
\end{eqnarray}
\begin{eqnarray}
\label{eq:newKG} 0=\frac{\xi \ddot {\phi} B'(\phi)}{H^2(1-6 \xi
B(\phi))}+\frac{3 \xi \dot {\phi} B'(\phi)}{H(1-6 \xi B(\phi))}
+\frac{R}{H^2}\frac{\xi^{2}{B'}^2(\phi)}{2(1-6 \xi B(\phi))}
+\frac{V'(\phi)\xi B'(\phi)}{H^2(1-6 \xi B(\phi))}.
\end{eqnarray}
We use the following dimensionless parameters to cast above equations in an autonomous form.
\begin{eqnarray}
\label{eq:omega} x &=& \frac{{\dot {\phi}}^{2}}{H^2(1-6 \xi
B(\phi))}, ~~y = \frac{2 V(\phi)}{H^2(1-6 \xi B(\phi))},~~
z = \frac{6\xi \dot {\phi} B'(\phi)}{H(1-6 \xi B(\phi))},\nonumber\\
\Omega &=& \frac{2 \rho}{H^2(1-6 \xi B(\phi))},~~ A=\frac{B'(\phi)\phi}{(1-6 \xi B(\phi))},~~ b=\frac{B''(\phi)\phi}{B'(\phi)},~~ c=\frac{V'(\phi)\phi}{V(\phi)},
\end{eqnarray}
where prime ($'$) and dot ($.$) represent the derivative with respect to $\phi$ and time, respectively.
Hence, the evolution equations (\ref{eq:newFried1}), (\ref{eq:newFried2}) and
(\ref{eq:newKG}) can be written as
\begin{eqnarray}
\label{eq:autonomous4}
\frac{dx}{d\ln a} &=& x' = 12X\frac{x}{z}-2x(\frac{Y}{6}-2)+xz,\nonumber\\
\frac{dy}{d\ln a} &=& y' = \frac{yz}{6\xi}\frac{c}{A}-2y(\frac{Y}{6}-2)+yz,\nonumber\\
\frac{dz}{d\ln a} &=& z' = 6X+\frac{z^2}{6\xi}\frac{b}{A}-z(\frac{Y}{6}-2)+z^2,\nonumber\\
\frac{dA}{d\ln a} &=& A' = \frac{z}{6\xi}(b+1)+Az,\nonumber\\
\frac{d\Omega}{d\ln a} &=& {\Omega}^{'} = \Omega(-3-3\omega-2(\frac{Y}{6}-2)+z).
\end{eqnarray}
The higher order derivative terms  having $\dot H$  and
$\ddot {\phi}$ in the autonomous system are
\begin{eqnarray}
\label{eq:xyr} X\equiv\frac{\xi \ddot {\phi} B'(\phi)}{H^2(1-6 \xi
B(\phi))},~~~ Y\equiv\frac{R}{H^2}.
\end{eqnarray}
The expressions of $\Omega$, $X$ and $Y$ can also be expressed in terms of $x, y, z$, and
are written as
\begin{eqnarray}
\label{eq:capxy}
\Omega&=&1-x-y-z,\nonumber\\
X(x,y,z)&=&-\frac{z}{2}-\frac{z^2}{18(4x+z^2)}\left(-6x+12y+\frac{z^2
b}{2\xi A}+\frac{yc}{\xi A}\right.
 + 3 (1-x-y-z)(1-3\omega) \Big{)},\nonumber\\
Y(x,y,z)&=&\frac{4x}{4x+z^2}\left(-6x+12y+\frac{z^2}{4\xi A}
\left(2b-\frac{yc}{x}\right)\right. + 3 (1-x-y-z)(1-3\omega)\Big{)}.
\end{eqnarray}

The energy density, pressure and the equation of state for a NMC scalar field model are defined as
\begin{eqnarray}
\label{eq:rhophi}
\rho_{\phi}&=&\frac{1}{2}{\dot {\phi}}^{2}+V(\phi)+3\xi( H \dot {\phi} B'(\phi)+H^{2} B(\phi)),\\
p_{\phi}&=&\frac{1}{2}{\dot {\phi}}^{2}-V(\phi)-\xi\left(2 H \dot
{\phi} B'(\phi)+ \dot {\phi}^{2} B''(\phi)+\ddot {\phi}
B'(\phi)+(2\dot{H}+3H^{2}) B(\phi)\right),\\
w_{\phi} &=& \frac{p_{\phi}}{\rho_{\phi}}= \frac{x-y-z-4 \xi x-2 X-2 \xi A (Y/6-1/2)}{x+y+z+3 \xi A}.
\end{eqnarray}
 
We shall work with the equations of autonomous system (\ref{eq:autonomous4}) to find the fixed points. We are  interested in stable solutions that can provide the late time cosmic acceleration. Therefore, we shall choose particular functional forms of $B(\phi)$ and $V(\phi)$ in the following section.

\begin{figure}[tbp]
\begin{center}
\begin{tabular}{c}
{\includegraphics[width=3in,height=2.5in,angle=0]{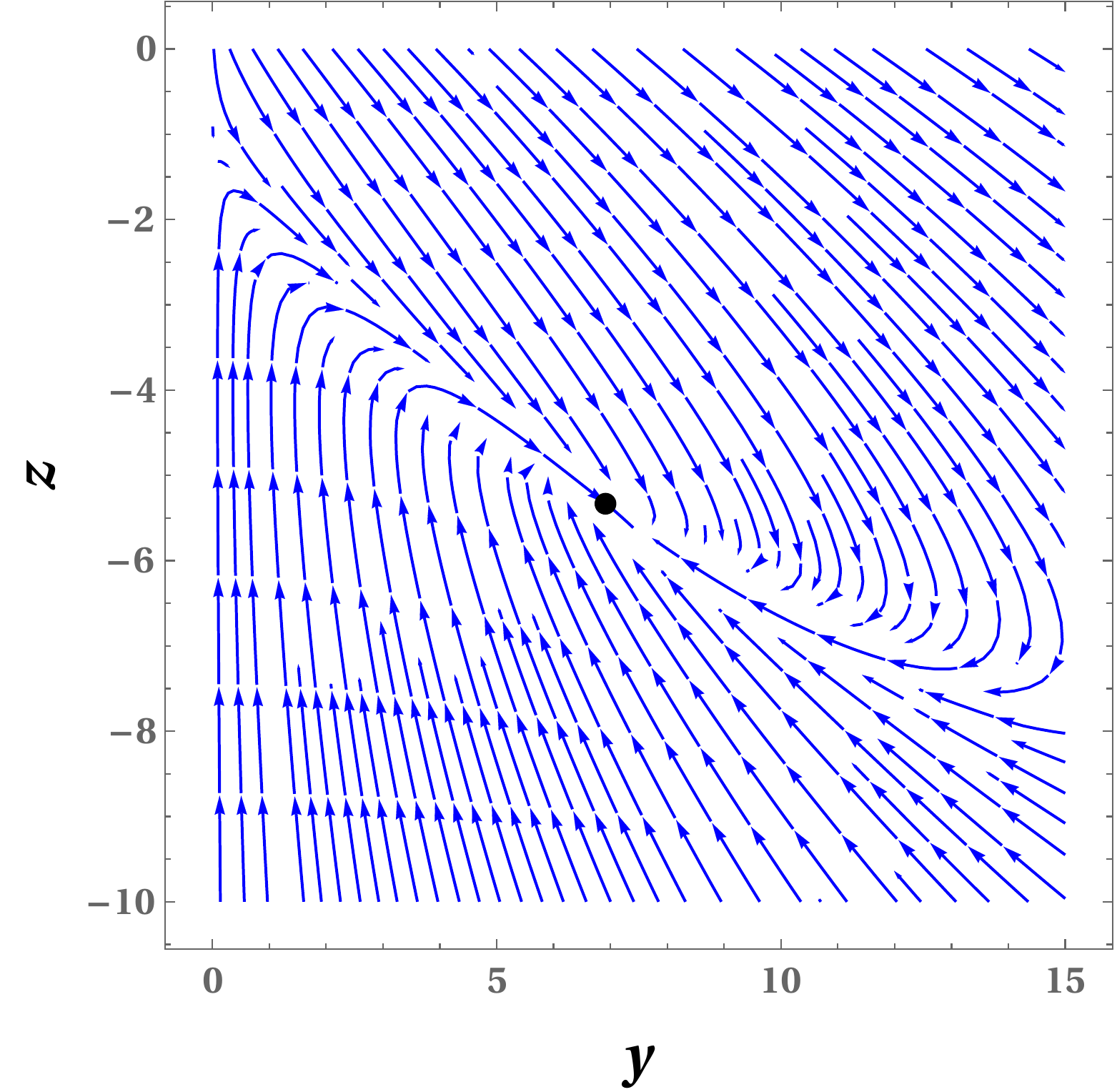}} 
\end{tabular}
\end{center}
\caption{The figure shows the stable fixed point for point 1 of model 1 with $\xi=2$ and $w=0$. The corresponding eigenvalues under the chosen parameters are $\mu_1=-5.333$, $\mu_2=-3$ and $\mu_3=-5.666$ that represent a stable point which is an attractive node. The black dot corresponds to the stable attractor point. }
\label{fig:KKLT}
\end{figure}

\section{Phase space analysis: stationary points and their stability}
\label{sec:phase}
\subsubsection{Model 1: $B(\phi) \propto {\phi}^2$, $V(\phi) = V_0 \frac{\phi^n}{\phi^n+M^n}$}
\label{sec:model1}

For model 1, we consider the KKLT model \cite{KKLT}
\begin{eqnarray}
V(\phi) = V_0 \frac{\phi^n}{\phi^n+M^n}
\label{eq:model1}
\end{eqnarray}
where $M$ and $n$ are free parameters. For the sake of simplicity, we set $n=2$ and $M=1/2$ \cite{KKLT}.

In these particular forms of $B(\phi)$ and $V(\phi)$,
\begin{eqnarray}
\label{eq:b}
b &=& \frac{B''(\phi)\phi}{B'(\phi)}=1, \\
c &=& \frac{V'(\phi)\phi}{V(\phi)}=\frac{2(1+3 \xi A)}{1+3 \xi A+2A}.
\label{eq:c1}
\end{eqnarray}
The following relation exists between $x$ and $z$ for $b=1$, 
\begin{eqnarray}
\label{eq:b1} x&=&\frac{\dot{\phi^2}}{H^2(1-6\xi B(\phi))}
=\frac{\dot{\phi^2}}{H^2(1-6\xi B(\phi))}\frac{{(6\xi
B'(\phi))}^2}{{(6\xi B'(\phi))}^2} \frac{\phi(1-6\xi
B(\phi))}{\phi(1-6\xi B(\phi))}=\frac{z^2}{72\xi^2  A}.
\end{eqnarray}

We put equation (\ref{eq:b1}) and $b=1$ in  equation (\ref{eq:capxy}), and find
 \begin{eqnarray}
\label{eq:b3}
\Omega&=&1-\frac{z^2}{72{\xi}^2 A}-y-z,\nonumber\\
X&=&-\frac{z}{2}-\frac{1}{1+18\xi^2 A}\left (\frac{z^2}{12}(6\xi -1)+y\xi (12\xi A+c) +3\xi^2 A \Omega (1-3\omega)\right ),\nonumber\\
Y&=&\frac{1}{1+18\xi^2  A}\left (\frac{z^2}{12\xi^2 A}(6\xi -1)+6y (2-3 c\xi )+3 \Omega (1-3\omega)\right).
\end{eqnarray}
Substituting equations (\ref{eq:b1}) and (\ref{eq:b3}) in (\ref{eq:autonomous4}), we finally obtain 
\begin{eqnarray}
\label{eq:b4}
y'&=&\frac{yz}{6\xi}\frac{c}{A}-2y\Big{(}\frac{1}{6(1+18\xi^2  A)}\left (\frac{z^2}{12\xi^2 A}(6\xi -1)+6y
 (2-3 c\xi ) +3(1-\frac{z^2}{72\xi^2  A}-y-z)(1-3\omega)\right )-2\Big{)}
+yz, \nonumber \\
z'&=&\left( -3z-\frac{6}{1+18\xi^2  A}\Big{(}\frac{z^2}{12}(6\xi
-1)+y\xi (12\xi A+c)+3\xi^2 A(1-\frac{z^2}{72\xi^2
A}-y-z)(1-3\omega) \Big{)}\right)\nonumber\\
&& +\frac{z^2}{6\xi A}-z\left( \frac{1}{6(1+18\xi^2
A)}\Big{(}\frac{z^2}{12\xi^2 A}(6\xi -1)+6y (2-3 c\xi )
+3(1-\frac{z^2}{72\xi^2  A}-y-z)(1-3\omega) \Big{)}-2 \right)
+z^2,\nonumber\\
 A'&=&\frac{z}{3\xi}+Az.
\end{eqnarray}
We shall investigate the equations of autonomous system (\ref{eq:b4}), and find the stationary points by equating the left hand side of equation (\ref{eq:b4}) to zero. Numerically, their stability can be obtained by the sign of corresponding eigenvalues.

\begin{enumerate}
\item 

\begin{eqnarray}
\label{eq:M1pt1}
y&=& \frac{3-28 \xi+60 \xi^2}{3(1-2 \xi)^2}, \qquad z= \frac{8 \xi}{1-2 \xi}, \qquad A=-\frac{1}{3 \xi}, \qquad \Omega=0,
\end{eqnarray}
The corresponding eigenvalues are given by,
\begin{eqnarray}
{\mu}_1 &=& \frac{8 \xi}{1-2 \xi} ~~~~ < 0 \qquad \text{for} \qquad \xi \neq 1/2, \qquad -\infty \leq \xi <0 \qquad \text{and} \qquad 1/2 < \xi \leq \infty  \nonumber \\
 {\mu}_2 &=& -3 (1+w) <0 \qquad \text{for} \qquad 1+w>0   \nonumber \\
  {\mu}_3 &=& -\frac{3 -10 \xi}{1-2 \xi} <0 \qquad \text{for} \qquad \xi \neq 1/2, \qquad -\infty \leq \xi < 3/10 \qquad \text{and} \qquad 1/2 < \xi \leq \infty
\end{eqnarray}
By looking the eigenvalues, this point is stable for above mentioned conditions. One can get expression of $Y$ by using equation (\ref{eq:xyr}) as 
\begin{eqnarray}
Y&=&\frac{R}{H^2}=6\left(2+\frac{\dot{H}}{H^2}\right)=\frac{12}{1-2 \xi}
\label{eq:M1Y}
\end{eqnarray}
On integrating equation (\ref{eq:M1Y}), we find the expression of $a(t)$:
\begin{eqnarray}
\label{eq:aY}
a(t)&=&a_0 \mid t-t_0 \mid^{\frac{1}{2-\frac{Y}{6}}}
\end{eqnarray}
where $a_0$ and $t_0$ are integration constant. The expression of $a(t)$ for the stationary point under consideration is given by
\begin{eqnarray}
\label{eq:M1a}
a(t)&=&a_0 \mid t-t_0 \mid^{-\frac{1-2 \xi}{4 \xi}}
\end{eqnarray}
One can use the following combination of the dimensionless variables to obtain the expression of $\phi(t)$
\begin{equation}
\frac{z}{6\xi A}=\frac{\dot{\phi}}{\phi H} \equiv \beta
\label{eq:beta}
\end{equation}
which tells us that
\begin{eqnarray}
\label{eq:phiY}
\phi(t)&=& \phi_0 \mid t-t_0 \mid^{\frac{\beta}{2-\frac{Y}{6}}}
\end{eqnarray}
where $\phi_0$ and $t_0$ are integration constant. Finally, we have
\begin{eqnarray}
\label{eq:M1phi}
\phi(t)&=& \phi_0 \mid t-t_0 \mid
\end{eqnarray}
This point is stable that can be noticed by eigenvalues. The expressions of $a(t)$ and $\phi(t)$ have power law behaviors that do not satisfy the de-Sitter condition. 

\item
\begin{eqnarray}
\label{eq:M1pt2}
y&=&0, \qquad z=\frac{4 \xi (1-3w)}{1-w-4 \xi}, \qquad A=-\frac{1}{3 \xi}, \qquad \Omega= \frac{(1-6 \xi)(3 - 16 \xi +3w (w+ 8 \xi-2)}{3(1-w-4 \xi)^2},
\end{eqnarray}
The corresponding eigenvalues are given by,
\begin{eqnarray}
{\mu}_1 &=& 3(1+w) < 0 \qquad \text{for} \qquad 1 +w <0 \nonumber\\
 {\mu}_2 &=&\frac{4 \xi (1-3w)}{1-w-4 \xi} <0 \qquad \text{for} \qquad 4 \xi (1-3w) <0 \nonumber\\
  {\mu}_3 &=& \frac{3-6w+3w^2-16 \xi+24w \xi}{2(1-w-4 \xi)} <0 \qquad \text{for} \qquad 3-6w+3w^2-16 \xi+24w \xi <0
\end{eqnarray}
This point is stable under given conditions. For this stationary point, we have
\begin{eqnarray}
Y&=&\frac{3(1-w)(1-3w)}{1-w-4 \xi}
\label{eq:M1pt1Y}
\end{eqnarray}
The expression of $a(t)$ and $\phi(t)$ are given by
\begin{eqnarray}
\label{eq:M1pt1a}
a(t)&=&a_0 \mid t-t_0 \mid^{\frac{2(1-w-4 \xi)}{3-16 \xi-3w^2}}\\
\phi(t)&=& \phi_0 \mid t-t_0 \mid^{\frac{4 \xi(1-3w)}{3w^2+16 \xi -3}}
\label{eq:M1pt1phi}
\end{eqnarray}
where $a_0$, $\phi_0$ and $t_0$ are integration constant. The expressions of $a(t)$ and $\phi(t)$ provide power law solutions that do not satisfy the de-Sitter condition. 

\item
\begin{eqnarray}
\label{eq:M1pt3}
y&=&0, \qquad z=12 \xi-2 \sqrt{36 \xi^2-6 \xi}, \qquad A=-\frac{1}{3 \xi}, \qquad  \Omega = 0,
\end{eqnarray}
The corresponding eigenvalues are following, and show negativity for below conditions.
\begin{eqnarray}
{\mu}_1 &=& 12 \xi- 2  \sqrt{36 \xi^2-6 \xi}<0 \qquad \text{for} \qquad 2  \sqrt{36 \xi^2-6 \xi}>12 \xi \nonumber\\
 {\mu}_2 &=& 3-3w- 12 \xi+ 2  \sqrt{36 \xi^2-6 \xi}<0 \qquad \text{for} \qquad 3+2  \sqrt{36 \xi^2-6 \xi}<3w + 12 \xi \nonumber\\
  {\mu}_3 &=& 6-12 \xi+2   \sqrt{36 \xi^2-6 \xi}<0 \qquad \text{for} \qquad 6+6  \sqrt{36 \xi^2-6 \xi}<12 \xi \nonumber\\
\end{eqnarray}
In this case, $Y=-6+72 \xi- 12  \sqrt{36 \xi^2-6 \xi}$, and the expressions of $a(t)$ and $\phi(t)$ are found to be
\begin{eqnarray}
\label{eq:M1pt3a}
a(t)&=&a_0 \mid t-t_0 \mid^{\frac{1}{3-12\xi+2\sqrt{36 \xi^2-6 \xi}}}\\
\phi(t)&=& \phi_0 \mid t-t_0 \mid^{-\frac{2 \xi}{2\xi+\sqrt{36 \xi^2-6 \xi}}}
\label{eq:M1pt3phi}
\end{eqnarray}
Again expressions of $a(t)$ and $\phi(t)$ provide power law solutions that do not qualify the de-Sitter condition.

\end{enumerate}

\begin{figure}[tbp]
\begin{center}
\begin{tabular}{c}
{\includegraphics[width=3in,height=2.5in,angle=0]{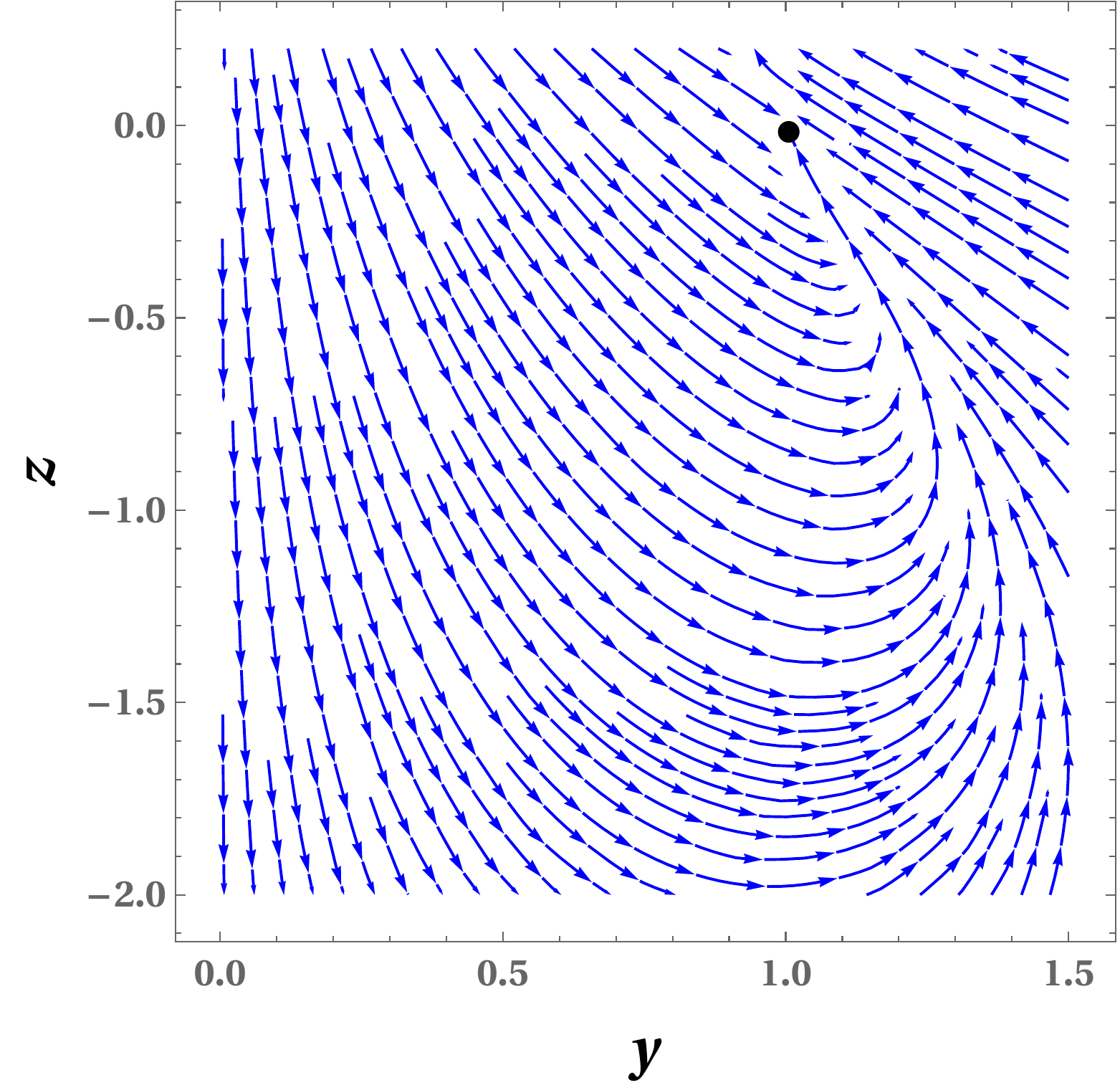}} 
\end{tabular}
\end{center}
\caption{The figure represents the phase portrait for point 2 of model 2 with $\xi=2$, $w=-1.01$ and $\sigma=10^{-16}$. The eigenvalues that correspond to these parameters are $\mu_1=-0.015$, $\mu_2=-2.992$ and $\mu_3=-0.030$. The black dot shows the stable attractor point which is an attractive node. }
\label{fig:Higgs}
\end{figure}

The phase portrait for the stable point 1 [Eq. (\ref{eq:M1pt1})] with $\xi=2$ and $w=0$ are shown in Fig. \ref{fig:KKLT}, where eigenvalues are $\mu_1=-5.333$, $\mu_2=-3$ and $\mu_3=-5.666$. By looking the eigenvalues, one can notice that the stable point is an  attractive node. All trajectories of phase portrait move towards the stable attractor point. 

\subsubsection{Model 2: $B(\phi) \propto {\phi}^2$, $V(\phi) = \frac{\lambda}{4} (\phi^2 - \sigma^2)^2$}
\label{sec:model2}
For model 2, we consider the Higgs potential \cite{Higgs}
\begin{eqnarray}
V(\phi) = \frac{\lambda}{4} (\phi^2 - \sigma^2)^2
\label{eq:model2}
\end{eqnarray}
Equations (\ref{eq:b3}) and (\ref{eq:b4}) will remain same except the dimensionless parameter $c$. Therefore, the expression of $c$ is given as
\begin{eqnarray}
c = \frac{4 A}{A- \sigma^2(2+6 \xi A)}
\label{eq:c}
\end{eqnarray}

In this model, we find following stationary point.

\begin{enumerate}
\item 
\begin{eqnarray}
\label{eq:M2pt1}
y&=&1, \qquad z=0, \qquad A=-\frac{1}{3 \xi}, \qquad \Omega=0,
\end{eqnarray}
The corresponding eigenvalues are given by,
\begin{eqnarray}
{\mu}_1 &=& 0,\nonumber \\
 {\mu}_2 &=&-3,\nonumber \\
  {\mu}_3 &=&-3(1+w).
\label{eq:M2pt1eigen}
\end{eqnarray}
This is not a stable point in the usual sense as one of the eigenvalue is zero. For this stationay point, we have
\begin{eqnarray}
Y&=&\frac{R}{H^2}=6\left(2+\frac{\dot{H}}{H^2}\right)=12\nonumber
\end{eqnarray}
which tells us that 
\begin{eqnarray}
\frac{\dot{H}}{H^2}=0,
\end{eqnarray}
and finally, we have
\begin{eqnarray}
\label{eq:a}
a(t)&=&a_0 e^{H_0(t-t_0)}
\end{eqnarray}
For this stationary point, $z=0$ which implies that $\dot{\phi}=0$ from equation (\ref{eq:beta}), and hence $\phi=\phi_0$ (constant). This point exhibits the de-Sitter behavior as one can notice that $\dot{H}=\dot{\phi}=0$. Though, it does not reflect a stable de-Sitter solution in the usual sense after looking the eigenvalues.

\item
\begin{eqnarray}
\label{eq:M2pt2}
y&=& \frac{3w^2+20 \xi-12 w \xi-3}{32 \xi}, \qquad z= \frac{3(1+w)}{2}, \qquad A=-\frac{1}{3 \xi}, \qquad \Omega= \frac{3(1+w)(1-6 \xi)}{16 \xi},
\end{eqnarray}
The corresponding eigenvalues are given by,
\begin{eqnarray}
{\mu}_1 &=& \frac{3(1+w)}{2} < 0 \qquad~~~~ \text{for} \qquad 1+w < 0,\nonumber \\
 {\mu}_2 &=& \frac{9w}{8}- \frac{3}{8}+ \delta_1 < 0 \qquad \text{for}\qquad \delta_1 <  \frac{3}{8}-\frac{9w}{8},\nonumber 
 \\
  {\mu}_3 &=& \frac{9w}{8}- \frac{3}{8}- \delta_1 < 0 \qquad \text{for} \qquad \delta_1 > \frac{9w}{8}- \frac{3}{8},
\end{eqnarray}
where
\begin{eqnarray}
\delta_1 &=& \frac{3 \sqrt{(1-6 \xi)^2 \xi^3 (6(w-1)(w+1)^2+(41+5(2-3w)w)\xi)}}{8 \xi^2 (1-6 \xi)}.
\label{eq:del1}
\end{eqnarray}
We have, $Y=\frac{15-9w}{2}$ in this case. The expressions of $a(t)$ and $\phi(t)$ are given by
\begin{eqnarray}
\label{eq:aMod2pt2}
a(t)&=&a_0 \mid t-t_0 \mid^{\frac{4}{3+3w}}\\
\phi(t)&=& \phi_0 \mid t-t_0 \mid^{-1}
\label{eq:phipoint2}
\end{eqnarray}
We see that $a(t)$ and $\phi(t)$ exhibit the power law behaviors. Therefore, it does not qualify the de-Sitter solution.

\item
\begin{eqnarray}
\label{eq:M2pt3}
y&=& 0, \qquad z= \frac{4 \xi(1-3w)}{1-w-4 \xi}, \qquad A=-\frac{1}{3 \xi}, \qquad \Omega= \frac{(1-6 \xi)(3-16 \xi+3w(w+8 \xi-2))}{3(1-w-4 \xi)^2},
\end{eqnarray}
The corresponding eigenvalues are given by,
\begin{eqnarray}
{\mu}_1 &=& \frac{3-3w^2-20 \xi+12w \xi}{1-w-4 \xi} < 0 \qquad \text{for} \qquad 3w^2+20 \xi > 3+ 12w \xi < 0,\nonumber \\
 {\mu}_2 &=& \frac{4 \xi (1-3w)}{1-w-4 \xi}< 0 \qquad \text{for} \qquad \xi<0 \qquad \text{and} \qquad 1-3w<0, \nonumber \\
  {\mu}_3 &=& \frac{6w-3w^2+16 \xi-24w \xi-3}{2(1-w-4 \xi)}< 0 \qquad \text{for} \qquad 3w^2+24w \xi+3 > 6w+16 \xi.
\end{eqnarray}

In this case, $Y=\frac{3(1-w)(1-3w)}{1-w-4 \xi}$. The expressions of $a(t)$ and $\phi(t)$ are given as
\begin{eqnarray}
\label{eq:aMod2pt3}
a(t)&=&a_0 \mid t-t_0 \mid^{\frac{2(1-w-4 \xi)}{3-3w^2-16 \xi}}\\
\phi(t)&=& \phi_0 \mid t-t_0 \mid^{\frac{4 \xi(1-3w)}{3w^2+16 \xi-3}}
\label{eq:phipoint3}
\end{eqnarray}

Again, by looking $a(t)$ and $\phi(t)$, this point does not provide the de-Sitter solution.

\end{enumerate}

\begin{figure}[tbp]
\begin{center}
\begin{tabular}{c}
{\includegraphics[width=3in,height=2.5in,angle=0]{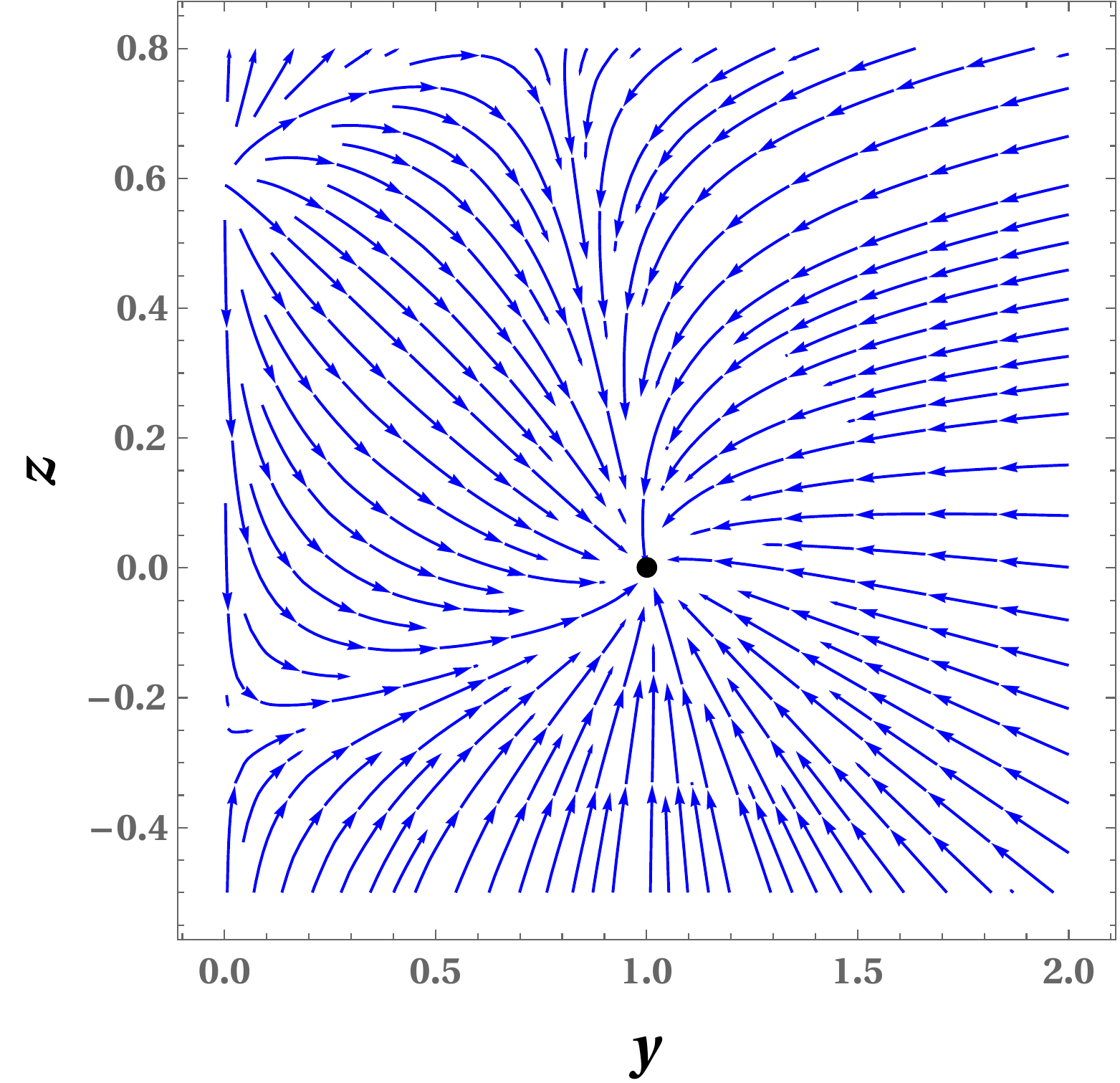}} 
\end{tabular}
\end{center}
\caption{The figure exhibits the phase space trajectories for point 3 of model 3 with $\xi=1/2$ and $w=0$. The stable fixed point is an attractive node that has eigenvalues $\mu_1=-3$, $\mu_2=-1.5-2.51i$ and $\mu_3=-1.5+2.51i$. }
\label{fig:inv}
\end{figure}

In Fig. \ref{fig:Higgs}, we display the phase space trajectories for point 2 [Eq. (\ref{eq:M2pt2})] with $\xi=2$,  $w=-1.01$ and $\sigma=10^{-16}$. The eigenvalues under the chosen parameters are $\mu_1=-0.015$, $\mu_2=-2.992$ and $\mu_3=-0.030$. The stable point is an  attractive node.  

\subsubsection{Model 3: $B(\phi) \propto {\phi}^2$, $V(\phi) = \frac{V_0}{\phi}$}
\label{sec:model3}
For model 3, we consider the inverse potential,
\begin{eqnarray}
V(\phi) = \frac{V_0}{\phi}
\label{eq:model3}
\end{eqnarray}
In this case, $c = -1$, and the stationary point are as follows.

\begin{enumerate}
\item 
\begin{eqnarray}
\label{eq:M3pt1}
y&=& \frac{3w^2+10 \xi+18w \xi-3}{2 \xi}, \qquad z= -6 (1+w), \qquad A=-\frac{1}{3 \xi}, \qquad \Omega=2-3w+ \frac{3(1+w)}{\xi},
\end{eqnarray}
The corresponding eigenvalues are given by,
\begin{eqnarray}
{\mu}_1 &=& -6 (1+w) < 0  \qquad \text{for} \qquad 1+w>0,\nonumber \\
 {\mu}_2 &=& \frac{3 \xi^2(3+w-6\xi(3+w))-\sqrt{3} \delta_2}{4 \xi^2(6 \xi-1)} < 0  \qquad \text{for} \qquad \xi \neq 0, 1/6 ~ \text{and} ~ \sqrt{3} \delta_2 > 3 \xi^2(3+w-6\xi(3+w)),\nonumber \\
 {\mu}_3 &=& \frac{3 \xi^2(3+w-6\xi(3+w))+\sqrt{3} \delta_2}{4 \xi^2(6 \xi-1)}< 0  \qquad \text{for} \qquad \xi \neq 0, 1/6 ~ \text{and} ~ \sqrt{3} \delta_2 < 3 \xi^2(6\xi(3+w)-3-w),
\end{eqnarray}
where
\begin{eqnarray}
\delta_2 &=& \sqrt{\xi^3(6\xi-1)(72(1-w)(1+w)^2-3\xi(73+w(254+w(161-24w)))+2(\xi+15w \xi)^2)}.
\label{eq:del2}
\end{eqnarray}

In this case, $Y=-3(5+9w)$. The expressions of scale factor and scalar field are given by
\begin{eqnarray}
\label{eq:M3pt1a}
a(t)&=&a_0 \mid t-t_0 \mid^{\frac{2}{9+9w}}\\
\phi(t)&=& \phi_0 \mid t-t_0 \mid^{2/3}
\label{eq:M3pt1phi}
\end{eqnarray}
This point shows the power law behavior of $a(t)$ and $\phi(t)$ that does not satisfy the de-Sitter condition. 
 
\item
\begin{eqnarray}
\label{eq:M3pt2}
y&=& \frac{6+ \xi(66 \xi-47)}{6(1 -\xi)^2}, \qquad z= \frac{10 \xi}{1-\xi}, \qquad A=-\frac{1}{3 \xi}, \qquad \Omega= 0,
\end{eqnarray}
The corresponding eigenvalues are given by,
\begin{eqnarray}
{\mu}_1 &=& \frac{10 \xi}{1-\xi} < 0  \qquad \text{for} \qquad \xi < 0, \neq 1, \nonumber \\
 {\mu}_2 &=& \frac{11 \xi-6}{2(1-\xi)}< 0  \qquad \text{for} \qquad \xi <6/11, \neq 1, \nonumber \\
  {\mu}_3 &=& \frac{3w \xi-3w-2 \xi-3}{1- \xi}< 0  \qquad \text{for} \qquad \xi \neq 1 \qquad \text{and} \qquad 3w \xi < 3w +2 \xi+3.
\end{eqnarray}

In this case, $Y=\frac{45}{1-\xi}-33$, and we have 
\begin{eqnarray}
\label{eq:M3pt2a}
a(t)&=&a_0 \mid t-t_0 \mid^{\frac{2(\xi-1)}{15 \xi}}\\
\phi(t)&=& \phi_0 \mid t-t_0 \mid^{2/3}
\label{eq:M3pt2phi}
\end{eqnarray}
Similar to point 1 it does not qualify the de-Sitter condition.

\item 
\begin{eqnarray}
\label{eq:M3pt3}
y&=&1, \qquad z=0, \qquad A=\frac{1}{12 \xi}, \qquad \Omega=0,
\end{eqnarray}
The corresponding eigenvalues are given by,
\begin{eqnarray}
{\mu}_1 &=& -3(1+w) < 0 \qquad \text{for} \qquad1+ w>0, \nonumber \\
 {\mu}_2 &=&- \frac{6\xi+9\xi^2 \sqrt{3\xi^2(2+3\xi)(6-71\xi)}}{2\xi(2+3\xi)}< 0 \qquad \text{for} \qquad \xi \neq 0 \qquad \text{and} \qquad 6\xi+9\xi^2 \sqrt{3\xi^2(2+3\xi)(6-71\xi)} > 0, \nonumber \\
  {\mu}_3 &=& \frac{\sqrt{3}\xi(6-71 \xi)}{2\sqrt{\xi^2(2+3\xi)(6-71\xi)}}-\frac{3}{2} < 0 \qquad \text{for} \qquad \xi \neq 0, 6/71 \qquad \text{and} \qquad \frac{\sqrt{3}\xi(6-71 \xi)}{2\sqrt{\xi^2(2+3\xi)(6-71\xi)}}<\frac{3}{2}.
\end{eqnarray}

In this case, $Y=12$ and the expression of scale factor can be obtained by using equation (\ref{eq:xyr}), and finally, we have
\begin{eqnarray}
\label{eq:aMod3pt3}
a(t)&=&a_0 e^{H_0(t-t_0)}
\end{eqnarray}
To get the expression of $\phi(t)$, we use equation (\ref{eq:beta}). But for underlying stationary point, $z=0$ which implies that $\dot{\phi}=0$, and hence $\phi=\phi_0$ (constant). Therefore, one can see that $\dot{H}=\dot{\phi}=0$ that reflects the stable de-Sitter solution.
\end{enumerate}

\begin{figure}[tbp]
\begin{center}
\begin{tabular}{c}
{\includegraphics[width=3in,height=2.5in,angle=0]{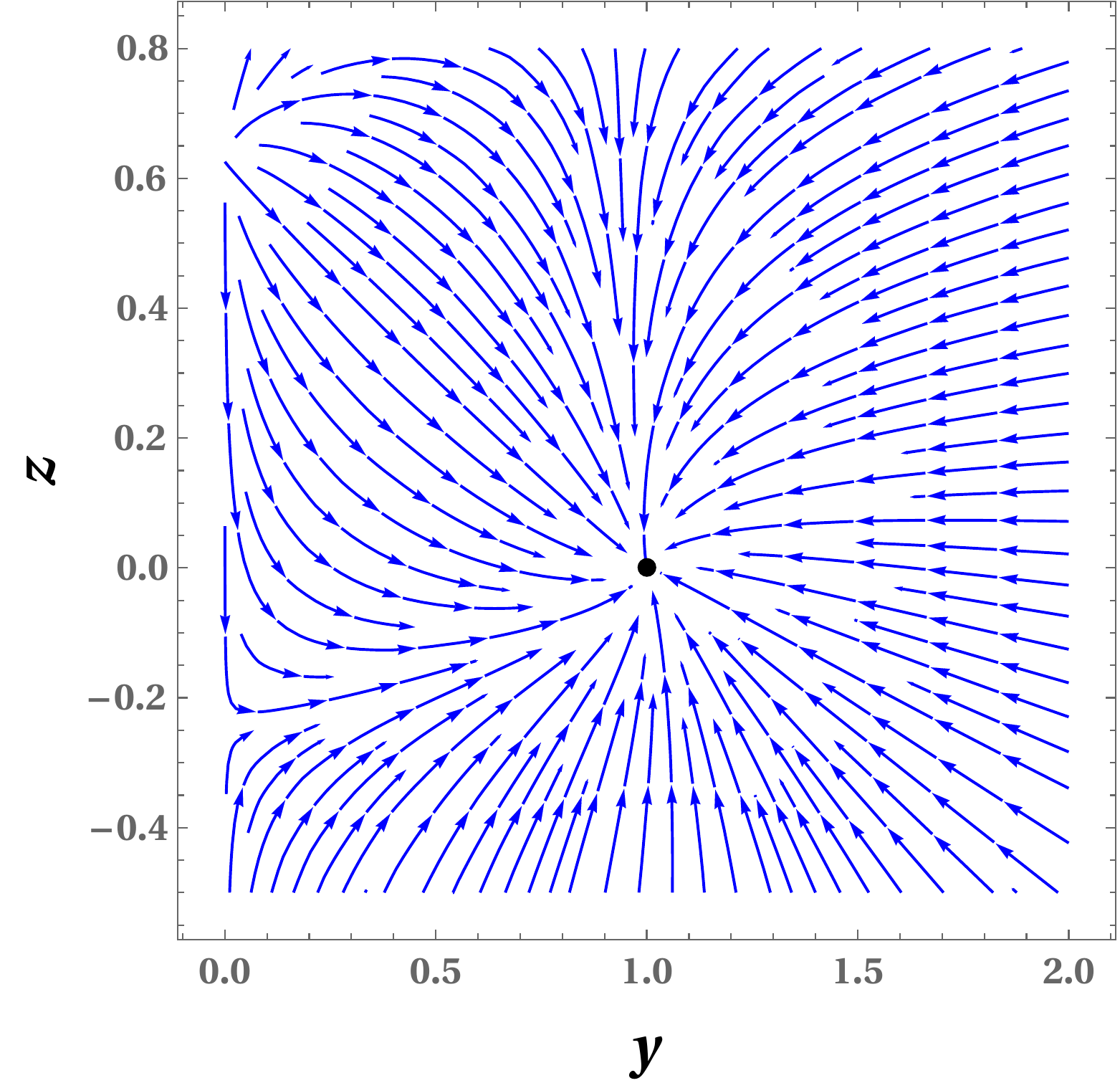}} 
\end{tabular}
\end{center}
\caption{The figure displays the phase portrait for point 3 of model 4. This is plotted for $\xi=1/5$ and $w=0$, and the eigenvalues are $\mu_1=-3$, $\mu_2=-1.5-1.5i$ and $\mu_3=-1.5+1.5i$. The point is stable and behaves as an attractive node. }
\label{fig:Invs}
\end{figure}

Fig. \ref{fig:inv} exhibits the phase portrait of point 3 [Eq.(\ref{eq:M3pt3})] with $\xi=1/2$ and  $w=0$. The eigenvalues for the chosen parameters are $\mu_1=-3$, $\mu_2=-1.5-2.51i$ and $\mu_3=-1.5+2.51i$. The point is a stable attractive node.

\subsubsection{Model 4: $B(\phi) \propto {\phi}^2$, $V(\phi) = \frac{V_0}{\phi^2}$}
\label{sec:model4}
For model 4, we consider the inverse square potential,
\begin{eqnarray}
V(\phi) = \frac{V_0}{\phi^2}
\label{eq:model4}
\end{eqnarray}
In this model, $c = -2$, and has following stationary point.

\begin{enumerate}
\item 
\begin{eqnarray}
\label{eq:mod3pt1}
y&=& \frac{3w^2+8 \xi+24 w \xi-3}{8 \xi}, \qquad z= -3 (1+w), \qquad A=-\frac{1}{3 \xi}, \qquad \Omega=3+ \frac{3(1+w)}{4\xi}
\end{eqnarray}
The corresponding eigenvalues are given by,
\begin{eqnarray}
{\mu}_1 &=& -3 (1+w) < 0  \qquad \text{for} \qquad 1+w>0, \nonumber\\
 {\mu}_2 &=& \frac{6 \xi^2(1-6\xi)-3\sqrt{2} \delta_3}{4 \xi^2(6 \xi-1)} < 0  \qquad \text{for} \qquad \xi \neq 0, 1/6 \qquad \text{and} \qquad 3\sqrt{2} \delta_3 > 6 \xi^2(1-6\xi),\nonumber \\
 {\mu}_3 &=& \frac{6 \xi^2(1-6\xi)+3\sqrt{2} \delta_3}{4 \xi^2(6 \xi-1)} < 0 \qquad \text{for} \qquad \xi \neq 0, 1/6 \qquad \text{and} \qquad 3\sqrt{2} \delta_3 + 6 \xi^2(1-6\xi) <0,
\end{eqnarray}
where
\begin{eqnarray}
\delta_3 &=& \sqrt{\xi^3(6\xi-1)(3(1-w)(1+w)^2-2\xi+4w \xi (8+9w)+4\xi^2(5+24w))}.
\label{eq:del2}
\end{eqnarray}

In this case, $Y=-6(1+3w)$, and the expressions of scale factor and field are given by
\begin{eqnarray}
\label{eq:M4pt1a}
a(t)&=&a_0 \mid t-t_0 \mid^{\frac{1}{3+3w}}\\
\phi(t)&=& \phi_0 \mid t-t_0 \mid^{1/2}
\label{eq:M4pt1phi}
\end{eqnarray}
It does not satisfy the de-Sitter condition.

\item 
\begin{eqnarray}
\label{eq:M4pt2}
y&=&1-6\xi, \qquad z=12\xi, \qquad A=-\frac{1}{3 \xi}, \qquad \Omega=0
\end{eqnarray}
The corresponding eigenvalues are given by,
\begin{eqnarray}
{\mu}_1 &=& -3, \nonumber \\
 {\mu}_2 &=&- 3 (1+w+4\xi) <0 \qquad \text{for} \qquad 1+w+4\xi <0,  \nonumber \\
  {\mu}_3 &=& 12\xi <0 \qquad \text{for} \qquad \xi <0.
\end{eqnarray}

We have, $Y=12+72\xi$, and the expressions of scale factor and field are
\begin{eqnarray}
\label{eq:M4pt2a}
a(t)&=&a_0 \mid t-t_0 \mid^{-\frac{1}{12\xi}}\\
\phi(t)&=& \phi_0 \mid t-t_0 \mid^{1/2}
\label{eq:M4pt2a}
\end{eqnarray}
It does not qualify the de-Sitter condition.

\item 
\begin{eqnarray}
\label{eq:M4pt3}
y&=&1, \qquad z=0, \qquad A=\frac{1}{6 \xi}, \qquad \Omega=0
\end{eqnarray}
The corresponding eigenvalues are given by,
\begin{eqnarray}
{\mu}_1 &=& -3(1+w) < 0 \qquad \text{for} \qquad1+ w>0, \nonumber \\
 {\mu}_2 &=&- \frac{3(\xi+3\xi^2 +\sqrt{\xi^2(1+3\xi)(1-13\xi)}}{2\xi(1+3\xi)}\nonumber\\
 && < 0 \qquad \text{for} \qquad \xi \neq 0,-1/3 \qquad \text{and} \qquad 3(\xi+3\xi^2 +\sqrt{\xi^2(1+3\xi)(1-13\xi)} > 0,\nonumber \\ 
{\mu}_3 &=& \frac{3 \sqrt{\xi^2-10\xi^3-39\xi^4}}{2\xi(1+3\xi)}-\frac{3}{2} < 0 \qquad \text{for} \qquad \xi \neq 0, -1/3 \qquad \text{and} \qquad \frac{3 \sqrt{\xi^2-10\xi^3-39\xi^4}}{2\xi(1+3\xi)} < \frac{3}{2}.
\end{eqnarray}

In this case, $Y=12$, and we have
\begin{eqnarray}
\label{eq:M4pt3a}
a(t)&=&a_0 e^{H_0(t-t_0)}
\end{eqnarray}
For this stationary point, $z=0$ which implies that $\dot{\phi}=0$, and hence $\phi=\phi_0$ (constant). Therefore, it shows the stable de-Sitter behavior. 

\end{enumerate}

Finally, we show the phase space trajectories for point 3 [Eq.(\ref{eq:M4pt3})] in Fig. \ref{fig:Invs} with $\xi=1/5$ and  $w=0$. The corresponding eigenvalues are $\mu_1=-3$, $\mu_2=-1.5-1.5i$ and $\mu_3=-1.5+1.5i$. This is a stable point and behaves as an attractive node. 

\section{Behavior of $G_{eff}$}
\label{sec:Geff}
The effective Newtonian gravitational constant $G_{eff}$ can be expressed in terms of dimensionless parameters
\begin{equation}
G_{eff}=\frac{\kappa}{8 \pi (1-\kappa \xi B(\phi))} \equiv \frac{\kappa(1+3 \xi A)}{8 \pi}
\label{eq:Geff}
\end{equation}

In the literature, it has been shown that a family of de-Sitter solutions can be found in the absence of curvature term $B(\phi)R $ in the action \cite{topo}. Our study exhibits that in the presence of curvature term $B(\phi)R $, we also get a true de-Sitter solution in case of inverse and inverse square potentials as $G_{eff}$ and $\phi$ are constant. The conditions of de-Sitter solutions are $\dot{H}=0$, $\phi=$ constant, $w_{\phi} \simeq -1$ and $G_{eff}>0$, which are satisfied in models 2, 3 and 4. Though, by looking the eigenvalues of model 2 [Eq.(\ref{eq:M2pt1})], it does not give a stable de-Sitter solution. Only points 3 [Eqs.(\ref{eq:M3pt3}), (\ref{eq:M4pt3})] of models 3 and 4 can give rise a true stable de-Sitter solution. However, these properties were missed in our previous study in case of $B(\phi) \propto \phi^N$ and $V(\phi) \propto  \phi^n$ by using the exactly same autonomous system \cite{alam2012}.

\begin{figure}[tbp]
\begin{center}
\begin{tabular}{ccc}
{\includegraphics[width=2.5in,height=2.5in,angle=0]{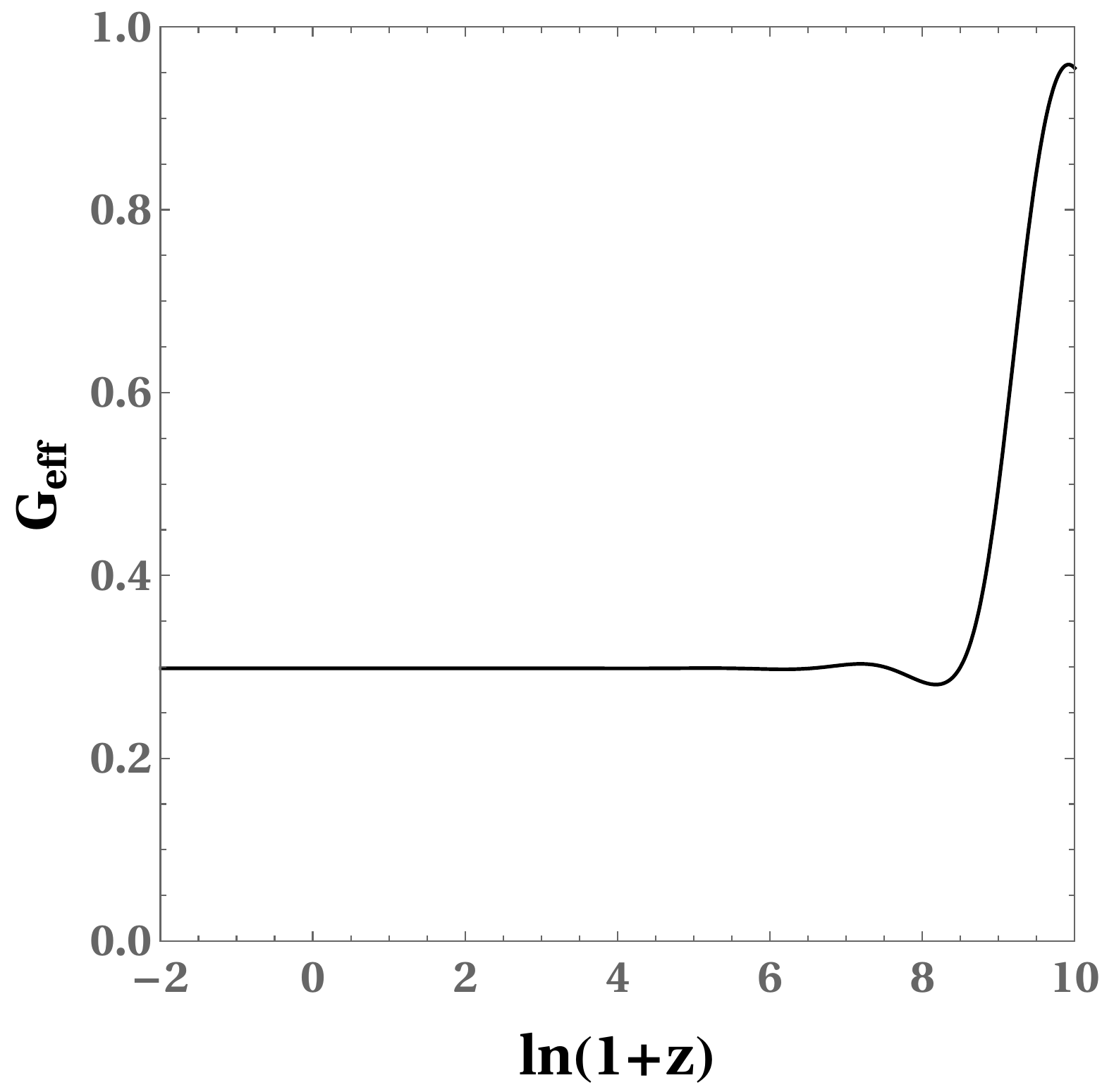}} &
{\includegraphics[width=2.5in,height=2.5in,angle=0]{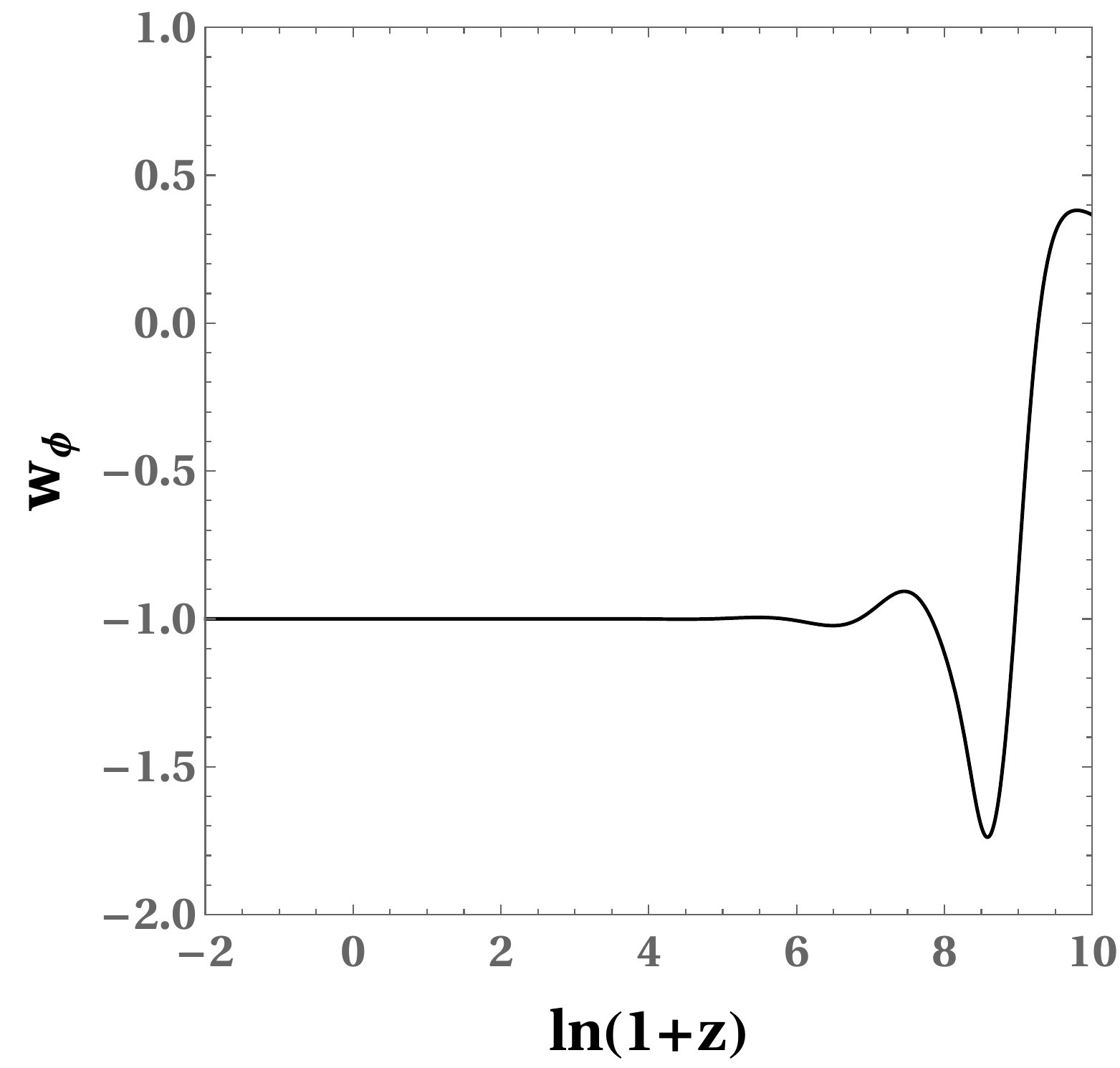}} 
\end{tabular}
\end{center}
\caption{ The figure exhibits the numerical evolution of $G_{eff}$ and $w_{\phi}$ versus redshift $z$ for model 3. We use $\xi=1$ and $w=0$ in the numerical evolution. The $G_{eff}$ remains positive during the entire evolution while $w_{\phi}$ passes through a short phantom phase in the past and moves towards $w_{\phi} \simeq -1$ around the current epoch that gives a stable de-Sitter solution. }
\label{fig:GN}
\end{figure}

The numerical evolution of $G_{eff}$ and $w_{\phi}$ versus redshift $z$ are depicted in Fig. \ref{fig:GN} for model 3. We find similar behavior of $G_{eff}$ and $w_{\phi}$ for other underlying models. In what follows, we restrict ourselves to model 3. The numerical evolution of $G_{eff}$ remains positive/constant for most of the period of evolution. Before moving towards the stable de-Sitter, $w_{\phi}$ pass through a short phantom phase in the past, and at late time it provides an observed value $w_{\phi} \simeq -1$ with $G_{eff}>0$.

\section{Conclusion}
\label{sec:conc}
In this paper, we considered four different potentials such as KKLT, Higgs, inverse and inverse square. We have investigated the phase space analysis for four NMC scalar field models having $F(\phi)R= 1-\xi\phi^2 $, by using a specific set of dimensionless variables. We chose the same equations of autonomous system that has been used in our previous paper \cite{alam2012}. Since, in the previous study, we missed important properties of de-Sitter solutions. For instance, we captured the de-Sitter solution and transient phase of dark energy with negative effective gravitational constant ($G_{eff}<0$), and if this is possible then the universe would be different from our real universe. Therefore, we re-investigated the dynamical behavior of NMC scalar field model with four different potentials. In the present study, we used the same autonomous system with $B(\phi) \propto \phi^2$ and KKLT, Higgs, inverse and inverse square potentials. We know that a de-Sitter solution can be obtained if $\dot{H}=0$, $\phi=$constant, $w_{\phi}\simeq -1$ and $G_{eff}>0$. Conclusively, our study showed that a true de-Sitter solution can be found with positive effective gravitational constant ($G_{eff}>0$). In case of KKLT, we did not find a de-Sitter solution. However, in case of Higgs and models 3 and 4, the de-Sitter conditions were trivially satisfied but seeing the eigenvalues of Higgs potential, it did not provide a stable de-sitter solution in usual sense as one of the eigenvalue is zero, see equation (\ref{eq:M2pt1eigen}).

The phase portraits of stable points of models 1, 2, 3 and 4 were displayed in Figs. \ref{fig:KKLT}, \ref{fig:Higgs}, \ref{fig:inv} and \ref{fig:Invs}, respectively. All phase space trajectories around the present epoch converged to $w_{\phi}\simeq -1$, and moved towards the stable attractor point which appeared as an attractive node. The numerical evolution of  $G_{eff}$ and $w_{\phi}$ versus redshift $z$ were shown in Fig. \ref{fig:GN}. The effective Newtonian gravitational constant is positive throughout the evolution. The evolution of $w_{\phi}$ exhibited the transient phantom phase in the past and late time acceleration around the present epoch before approaching the stable attractor point. 

\section*{Acknowledgments}
The work is partially financially  supported  by the Ministry of Education and Science of the Republic of Kazakhstan, Grant No. AP08052197.

\end{document}